\documentclass[apj]{emulateapj}
\begin{document}

\slugcomment{Accepted for publication in ApJ Letters}
\title{Hubble Space Telescope Images of Red
Mergers: \\ How Dry are They?}
\author{Katherine E. Whitaker\altaffilmark{1} \& Pieter G. van Dokkum\altaffilmark{1}}
\altaffiltext{1}{Department of Astronomy, Yale University, P.O. Box 208101, New Haven, CT 06520}
\email{katherine.whitaker@yale.edu}

\begin{abstract}

Mergers between red galaxies are observed to be common in the nearby
Universe, and are thought to be the dominant mechanism by which
massive galaxies grow their mass at late times.
These ``dry'' mergers can be readily identified in very deep ground
based images, thanks
to their extended low surface brightness tidal features.
However, ground-based
images lack the required resolution to determine the morphologies
of the merging galaxies, and to measure the amount of dust and
associated gas. We present HST/ACS
and WFPC2 observations of a sample of 31 bulge-dominated red-sequence galaxies 
at $z\sim 0.1$, comprised of ongoing mergers, merger remnants, and 
undisturbed galaxies.  Nearly all galaxies have early-type morphologies and
most are well-fit by r$^{1/4}$ law surface brightness profiles.
We find that only 10\% of the galaxies show evidence for the presence of dust. 
The amount of cold gas (or its upper limit)
is calculated from the mean color-excess, assuming 
a simple relation between gas mass and dust mass.   
The gas mass is low for all galaxies, and we find that
M$_{\mathrm{gas}}$/M$_{\mathrm{stellar}} \lesssim 3\times10^{-4}$.
We infer that red mergers in the nearby Universe mostly involve
early-type galaxies containing little cold gas and dust.
This may imply that the progenitors were mostly devoid of gas and/or
that feedback mechanisms are very effective in preventing the gas to
cool. The lack of gas in these objects
may also imply a relatively large fraction of binary
black holes in the centers of massive ellipticals.

\end{abstract}


\section{Introduction}

In currently popular models massive
elliptical galaxies are expected to have assembled a significant fraction
of their 
final mass at redshifts $z<1$ through mergers (e.g., de Lucia et al.\ 2006).
As elliptical galaxies follow tight scaling relations and have
old stellar populations, these mergers cannot involve large amounts
of gas and associated star formation. Instead,
recent studies have advanced the idea of the continued assembly of
elliptical galaxies through mergers between galaxies which are
already on the red sequence
(e.g., van Dokkum et al.\ 1999,
Bell et al. 2004, 2006, van Dokkum 2005 [hereafter, vD05],
Tran et al.\ 2005, White et al. 2007, 
Faber et al. 2007,  Zheng et al. 2007, McIntosh et al. 2007).
These so-called ``dry'' mergers are now thought to be the dominant mode
of growth of massive ($>$ several M$_{\star}$) galaxies at $0<z<1$,
building up the high-mass end of the mass function 
but not changing the overall mass density of ellipticals.

It is not yet clear how much gas is involved in these mergers, that
is, how dry they are. The amount of gas provides an important constraint
on the efficiency of feedback in galaxy formation models
(see, e.g., Croton et al.\ 2006), and may also determine the
fate of the supermassive black holes of the progenitor galaxies
(e.g., Escala et al.\ 2004).
The red colors of massive mergers suggest that their light is
dominated by old stellar populations, but may also reflect the
presence of dust. Furthermore, we have only very limited information
on the morphologies of galaxies involved in dry mergers.  
Recent simulations by Feldmann et al.\ (2008) 
have demonstrated that some 
tidal features observed by vD05 are not necessarily the result of
spheroid-spheroid mergers, but could also be
kinematically cold material from the merger of a spheroid and
a disk-dominated system.  

Deep ground-based images are excellent for detecting low
surface-brightness features, but their limited resolution 
makes it difficult to address these questions.
The ongoing mergers and merger remnants identified in
vD05 have redshifts of $z\sim 0.1$, and
are barely resolved within the effective radius
(assuming $r_e \approx 3$\,kpc and $\approx 1\arcsec$ seeing).
Here we present {\em Hubble Space Telescope} ({\em HST}) images
with the Advanced Camera for Surveys (ACS) and the Wide Field
and Planetary Camera 2 (WFPC2) for a representative subset of
galaxies drawn from the vD05 sample of luminous
red galaxies. The resolution offered by HST allows us to
determine the morphologies of the red mergers and their
remnants, and to determine the amount of dust associated with
these mergers.

%=== Fig 1
\begin{figure*} %[htp]
\begin{center}
%\leavevmode
\epsscale{0.95}
\plotone{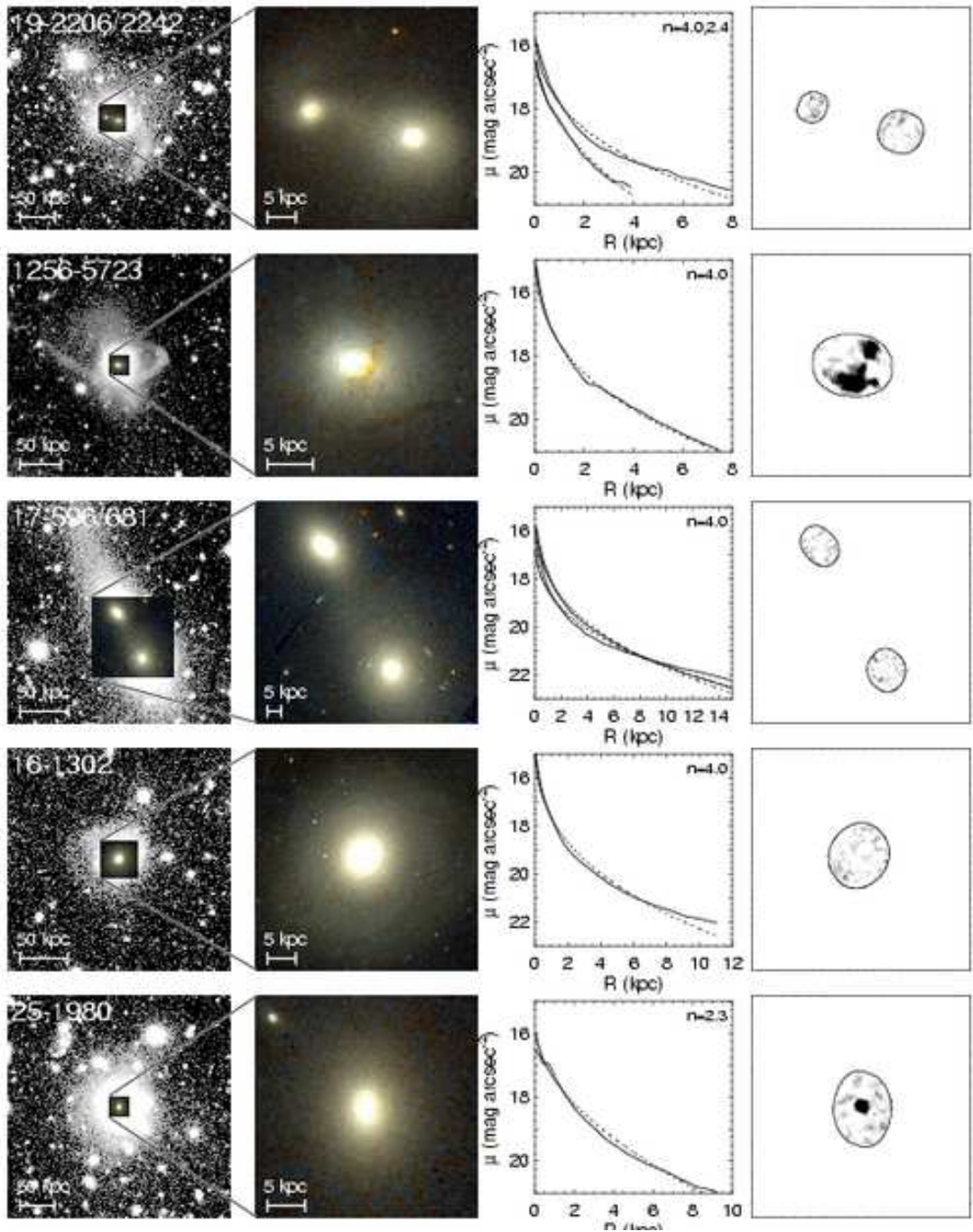}
\end{center}
\caption{Seven examples of galaxies from our sample: two ongoing mergers
and three objects with strong or weak tidal features (from
vD05). The left panels show very deep ground-based images
used to identify the tidal features in vD05. The second column shows
color images generated from the $V_{606}$ and $I_{814}$ HST data. Also shown
are surface brightness profiles (solid lines) with the best-fit Sersic
profiles overplotted (broken lines), and color excess images, where
black implies an excess of red light (i.e., dust).}
\label{hst}
\end{figure*}

\section{Data and Reduction}

The sample consists of 31 galaxies, 
selected out of a sample of 86 bright bulge-dominated, red galaxies at a 
redshift of z$\sim$0.1 discussed by vD05.  According to the visual tidal 
classification in vD05, the sample contains 3 undisturbed galaxies, 
5 weakly disturbed remnants, 5 strongly 
disturbed remnants, and 9 ongoing merging pairs.  
Each target was observed for one orbit, split in two exposures in the
$V_{606}$ filter and two exposures in the $I_{814}$ filters.
Fourteen galaxies were observed with
ACS, using the Wide Field Channel (WFC).
The instrument stopped operating due to an electrical short in January 2007,
so the remaining 17 galaxies were observed with the Wide
Field and Planetary Camera 2 (WFPC2). 

The targets observed by ACS have two 512\,s dithered 
exposures in each filter, enabling cosmic-ray detection.  
The ACS calibration pipeline performs the basic
image reduction and removal of cosmic-rays, as well as removing the geometric
distortions and combining the dithered data.  
The amplifiers on either side of the
chip resulted in two different background values in the pipeline product.  This
artifact was removed by subtracting the median background value for 
each amplifier.   

The targets observed by WFPC2 have two 400\,s dithered exposures.  The 
IRAF, STSDAS tasks $warmpix$ and $crrej$ remove warm pixels and cosmic-rays to
improve the standard pipeline calibrated data.  
The task {\sc L.A.Cosmic} (van Dokkum 2001) is 
further used to improve the cosmic-ray removal.

\section{Morphologies}

The majority of the sample appear to be normal early-type galaxies with 
no visible dust signatures.  The galaxies generally have smooth surface 
brightness profiles with little isophotal twisting or changes in 
ellipticity.  Low surface-brightness tidal features are present in the 
ongoing mergers and disturbed galaxies, but not as prominent as 
in the deeper ground-based images presented in vD05. 
Fig. \ref{hst} contains examples of galaxies in the sample 
with both ground-based imaging and two-color HST images.  Galaxies
with striking tidal features in the ground-based data (e.g., 19-2206/2242
and 17-596/681) appear regular in the shallow, high-resolution HST images.

\begin{figure}[h]
\begin{center}
%\leavevmode
\epsscale{1.1}
\plotone{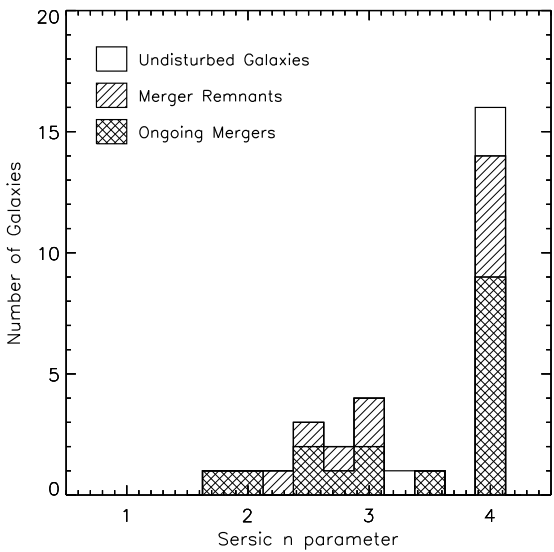}
\end{center}
\caption{The distribution
of the minimum $\chi^{2}$ best-fit Sersic $n$ parameter, for different
visual tidal classifications.
Undisturbed galaxies are open histograms, 
merger remnants
are diagonally hashed lines, and ongoing mergers are cross hashed lines.
Most galaxies appear to be bulge-dominated, and there is no correlation
between the extent of tidal features and the best-fitting Sersic
index.}
\label{sersic}
\end{figure}

To quantify the properties of the galaxies, we fit the Sersic (1968) 
r$^{1/n}$ law to the surface brightness profiles.  The free parameter $n$ 
is a quantitative measure of the bulge-to-disk ratio:
late-type spirals typically are dominated by exponential disks and
have $n\sim 1$, elliptical galaxies usually have
$n\sim 4$, and galaxies with a combination of varying
sized bulges and disks cover the middle ground. 
Figure \ref{sersic} shows the distribution of best-fitting Sersic parameters.
Most galaxies are well 
described with de Vaucouleurs profiles with $n\sim 4$, but not all. 
Different types of shading indicate differences in the presence
and extent of tidal features;
there is no correlation between the tidal features and the Sersic 
parameter. 

%=== Fig 2

\section{Deriving the Gas Mass}

The majority of the galaxies look like normal early-type galaxies, but a
few show small dips and other deviations in their surface brightness
profiles (e.g., 1256-5723 and
25-1980 in Fig. \ref{hst}). As the dips are more
pronounced in the $V_{606}$ band than in the $I_{814}$ band, the cause is
likely to be obscuration by dust. In the case of 1256-5723,
this is confirmed by the visual appearance of the galaxy: a
prominent dust lane winds across the galaxy in the HST image.
Assuming that reddening traces dust, the dust mass can
be constrained from the color-excess.  The color excess images 
are constructed as follows:

\begin{eqnarray}
E(V_{606}-I_{814}) & = & (V_{606}-I_{814})-(V_{606}-I_{814})_{\mathrm{model}} \\
       & = & -2.5 \log \left(\frac{[i(V_{606})/i(I_{814})]}{[i(V_{606})/i(I_{814})]_{\mathrm{model}}}\right),
\end{eqnarray}

\noindent where $i(V_{606})$ and $i(I_{814})$ are intensities and $V_{606}$
and $I_{814}$ are 
magnitudes.  The model $V_{606}-I_{814}$ color image contains no reddening 
due to dust, and is generated using an
ellipse fit 
with the visible dust regions masked. The color excess images are found in the 
last column in Fig. \ref{hst}, with darker regions signifying the
presence of dust. 

The reddening can be related to the mass in cold gas, assuming
that the dust traces cold gas, 
a  mass ratio of H to He of 3:1, and
an H gas-to-dust mass ratio of 100
(see, e.g., van Dokkum \& Franx 1995; Tomita et al.\ 2000):
\begin{equation}
M_{\mathrm{gas}} [M_{\odot}] = \Sigma \langle E_{V_{606}-I_{814}}\rangle
\Gamma_{V_{606}-I_{814}},
\end{equation}
where $\Sigma$ is the area of the feature (kpc$^{2}$), 
$\langle E_{V-I}\rangle$ is the mean reddening in that area, and 
$\Gamma_{V-I}$ is the mass reddening coefficient .
We adopt
$\Gamma_{V-I}\sim$5.6$\times$10$^{7}$ M$_{\odot}$ mag$^{-1}$ kpc$^{-2}$,
which was derived from the Galactic mass absorption coefficient in the
$V$ band \cite{Sadler85} and assuming that
$E(V_{606}-I_{814})_{z=0.1} = 0.38 A_{V, z=0}$ \cite{Rieke85,Cardelli89}.
The mean color excess value is measured only within the area
that is $>5\sigma$ above the 
background noise (shown as a black contour in the color excess images 
of Fig. \ref{hst}). 

Only 3 out of the 31 red galaxies (10\%) have significant detectable dust, 
with cold gas mass values ranging from 7.7-7.9 in log M$_{\odot}$.  
Of these 3 galaxies, one is an Sa galaxy with a striking dust lane that
was probably misclassified in vD05.  The other two galaxies are shown
in Fig. \ref{hst} (1256-5723 and 25-1980). 
The detection of dust in 1256-5723 is probably related to the 
startling tidal arm seen in ground-based images (see vD05, 
Feldmann et al. 2008).  This object may be an isolated example
of a merger with a galaxy with a prominent cold component.

For the remaining 28 galaxies (90\%)  we find no evidence for dust.
Upper limits on the dust masses in these galaxies were derived from the 
median absolute deviation of the distribution of mean color excess values.
To put the cold gas mass values of all 31 massive red galaxies in
context, we determine the gas-to-stellar mass ratio, using stellar 
mass values from catalogues by Kauffmann et al.\ (2003) and Gallazi 
et al.\ (2005). 
For those 11 galaxies missing from both catalogues,
the stellar masses were estimated from the empirical relationship 
between $R$-band magnitude (from vD05) and stellar mass
defined by the rest of the galaxies. The uncertainty in the masses
of these galaxies is $\sim 0.1$\,dex, smaller than the
uncertainty in the gas masses.

%=== Fig 3
\begin{figure}[h]
\begin{center}
%\leavevmode
\epsscale{1.1}
\plotone{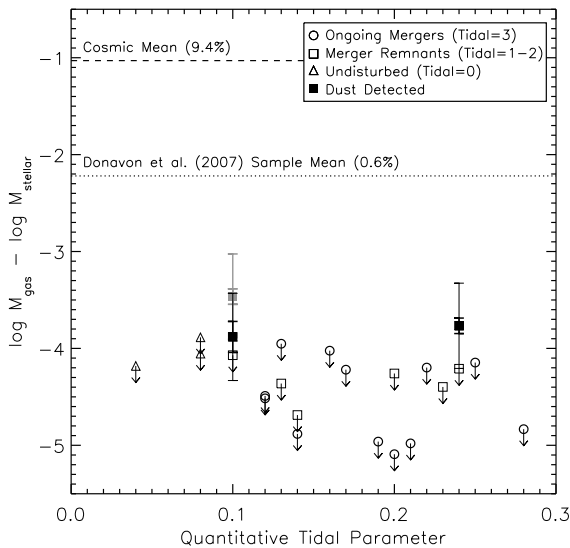}
\end{center}
\caption{Cold gas to stellar mass ratio in red galaxies 
with varying levels of  
tidal distortions (quantitative classification from vD05).  All galaxies have
M$_{\mathrm{gas}}$/M$_{\mathrm{stellar}} \lesssim 3\times10^{-4}$.  
Downward pointing
arrows indicate upper limits of the cold gas mass.  Those galaxies with 
significant dust detections are solid points (with the misclassified Sa
galaxy in gray), including random and systematic (larger) 
error bars.  
The cosmic mean (dashed line) and
the mean of the Donavon et al. sample (dotted line) are included for
reference. }
\label{ratio}
\end{figure}

Figure\ 3 shows the ratio of cold gas to stellar mass as a function of the 
quantitative tidal parameter $t$ defined in vD05.
This parameter describes the level of distortion by
measuring the median absolute deviation of the residuals from model fits.
The gas mass in these
galaxies is remarkable small, amounting to less
than $0.1$\,\% of the stellar mass even for the galaxies where dust
was detected.

\section{Discussion}

In this work, we use high-resolution images taken with
HST of red mergers identified in vD05
to determine the morphologies of red mergers 
and to measure the amount of dust and associated cold gas.
Nearly all galaxies have early-type 
morphologies and most are
well-fit by r$^{1/4}$ law surface brightness profiles. 
We find that 90\% of the galaxies exhibit
no visible dust signatures, while 10\% have detectable dust.  
The gas-to-stellar mass ratio is low for all galaxies, including
the three with detected dust, where 
M$_{\mathrm{gas}}$/M$_{\mathrm{stellar}} \lesssim 3\times10^{-4}$.

It is interesting to compare this result to those of
Donovan et al.\ (2007), who examined a sample of red elliptical
galaxies with significant amounts of neutral hydrogen (as traced
by 21\,cm observations).
The mean gas-to-stellar mass ratio of the Donovan et al.\
galaxies that would have
been selected by the vD05 criterion is 0.006, about an order of 
magnitude greater than the red ellipticals in this study.  
This implies that the H{\sc i}-rich systems studied by Donovan et
al.\ are not representative for the general population of
red mergers, that the dust is not a good tracer of the cold
gas in these systems, or that the cold gas typically resides at
large radii ($\gg 1 r_e$)
and hence would escape detection in our study.

The most straightforward interpretation of our results is that
red mergers in the nearby Universe mostly involve
early-type galaxies containing very little cold gas.
Elliptical galaxies generally have substantial
amounts of hot gas and it appears that this gas is not able
to condense and form stars, even during mergers.
Central active nuclei may play a role in preventing the gas from
cooling
(e.g., Croton et al.\ 2006), and it will be interesting to determine
whether the merger remnants show enhanced low-level AGN activity.

The absense of large amounts of gas in these 
objects\footnote{Note that almost all elliptical galaxies at $z\sim 0$
have trace amounts of gas and dust at levels substantially below 
our detection limit (van Dokkum \& Franx 1995; Tomita et al.\ 2000).}
has interesting
implications for the growth of supermassive black holes. It has
been argued that the massive gas disks observed in the centers
of gas-rich mergers may be instrumental in removing angular momentum
from the binary black holes that result from a major merger
(e.g., Escala et al.\ 2005). As pointed out by Hoffman \& Loeb
(2007), dry mergers of massive galaxies with little gas should
lead to many stable black hole binaries, and subsequently to
unstable triples. The lightest of the three black holes can then be
ejected, which has interesting consequences.
Our data lend indirect support to such scenarios.
A further possible implication is that the black hole merger rate,
which is relevant for gravitational wave detection experiments,
does not necessarily track the galaxy merger rate.

One of the main uncertainties in our study is the conversion from
reddening to gas mass. Deep H{\sc i} observations of the galaxies
in this sample could constrain the amount of cold gas more
directly, although the required integration times would be
very large. The amount of cold gas can also be constrained through
its effect on the star formation rate, which can be done by
analyzing the spectra
and spectral energy distributions of the galaxies (see Kaviraj
\& van Dokkum 2008).
Finally, it will be interesting to search for analogs
of these systems at even lower redshifts.

\acknowledgements{We thank the referee for helpful comments. Support
from NASA grant HST GO-10809.01-A is gratefully acknowledged.}

\end{document}